\documentclass[conference]{IEEEtran}
\IEEEoverridecommandlockouts
\usepackage{cite}
\usepackage{amsmath,amssymb,amsfonts}
\usepackage{makecell}
\usepackage{booktabs}
\usepackage{epsfig}
\usepackage{algorithmic}
\usepackage{graphicx}
\usepackage{textcomp}
\usepackage{xcolor}
\usepackage{multirow}
\def\BibTeX{{\rm B\kern-.05em{\sc i\kern-.025em b}\kern-.08em
    T\kern-.1667em\lower.7ex\hbox{E}\kern-.125emX}}

\makeatletter
\newcommand{\linebreakand}{%
  \end{@IEEEauthorhalign}
  \hfill\mbox{}\par
  \mbox{}\hfill\begin{@IEEEauthorhalign}
}
\makeatother

\begin{document}

\title{Frequency-regularized Neural Representation Method for Sparse-view Tomographic Reconstruction\\
\thanks{\textsuperscript{*}Corresponding author: Si Li. This work is supported by the Natural Science Foundation of Guangdong under 2022A1515012379. }
}

\author{\IEEEauthorblockN{Jingmou Xian}
\IEEEauthorblockA{\textit{School of Computer Science} \\
\textit{Guangdong University of Technology}\\
Guangzhou, China \\
2112205127@mail2.gdut.edu.cn}
\and

\IEEEauthorblockN{Jian Zhu}
\IEEEauthorblockA{\textit{School of Computer Science} \\
\textit{Guangdong University of Technology}\\
Guangzhou, China \\
dr.zhuj@gmail.com}
\and

\IEEEauthorblockN{Haolin Liao}
\IEEEauthorblockA{\textit{School of Computer Science} \\
\textit{Guangdong University of Technology}\\
Guangzhou, China \\
2112205270@mail2.gdut.edu.cn}

\linebreakand
\IEEEauthorblockN{Si Li\textsuperscript{*}}
\IEEEauthorblockA{\textit{School of Computer Science} \\
\textit{Guangdong University of Technology}\\
Guangzhou, China \\
sili@gdut.edu.cn}

}
\maketitle

\begin{abstract}
Sparse-view tomographic reconstruction is a pivotal direction for reducing radiation dose and augmenting clinical applicability. While many research works have proposed the reconstruction of tomographic images from sparse 2D projections, existing models tend to excessively focus on high-frequency information while overlooking low-frequency components within the sparse input images. This bias towards high-frequency information often leads to overfitting, particularly intense at edges and boundaries in the reconstructed slices. In this paper, we introduce the Frequency Regularized Neural Attenuation/Activity Field (Freq-NAF) for self-supervised sparse-view tomographic reconstruction. Freq-NAF mitigates overfitting by incorporating frequency regularization, directly controlling the visible frequency bands in the neural network input. This approach effectively balances high-frequency and low-frequency information. We conducted numerical experiments on CBCT and SPECT datasets, and our method demonstrates state-of-the-art accuracy.
\end{abstract}


\section{Introduction}
\label{sec:intro}

Medical imaging aims to reconstruct computed images from measurement data obtained through physical sensors to visualize the internal structure of living organisms \cite{ref1, ref2}. For instance, Cone-beam computed tomography (CBCT) and Single Photon Emission Computed Tomography (SPECT) require measurements of projection data. In practice, it is often necessary to reduce the number of measurements required to obtain high-quality medical images to minimize radiation exposure to patients. However, due to the loss of information during the sparse sampling measurement process, sparse-view image reconstruction becomes an ill-posed inverse problem. Sparse-view tomographic image reconstruction is designed to recover the volume attenuation or activity field from a limited number of projections. This is a challenging task because the scarcity of views can lead to noticeable artifacts. 
% In contrast, traditional CBCT typically involves hundreds of images, while the input images for sparse-view CBCT are about $ 10 \times$ fewer.

% Medical imaging aims to reconstruct computed images from measurement data acquired by physical sensors, enabling visualization of the internal structures of living organisms \cite{ref1, ref2}. For instance, in Cone-beam computed tomography (CBCT) and single-photon emission computed tomography (SPECT), projection data is measured. In practical applications, there is often a need to reduce the number of measurements required for reconstructing high-quality medical images. Specifically, sparse-view tomographic image reconstruction offers a solution for lowering the radiation exposure to patients. However, due to information loss during sparse sampling, sparse-view image reconstruction becomes an ill-posed inverse problem. Sparse-view tomographic image reconstruction aims to retrieve the volumetric attenuation or activity field from a limited number of projections. This is a challenging task, and insufficient views can lead to prominent artifacts. In comparison, the traditional CBCT captures hundreds of images. The inputs of sparse-view CBCT are $ 10 \times$ fewer. 

\begin{figure}[t]
\begin{minipage}[b]{1.0\linewidth}
\footnotesize
  \centering
    \quad NAF  \quad \quad \quad \quad \ \  \ \quad  Freq-NAF(Ours) \quad \quad \quad \quad \ \  Ground Truth
  % \centerline{\quad \quad \ \ NAF \quad \quad \ \  \ \  Ground Truth \quad \ Freq-NIF(Ours)}\medskip
% \centerline{\epsfig{figure=fig/motivation.eps,width=8.5cm}}
\includegraphics[width=8.5cm]{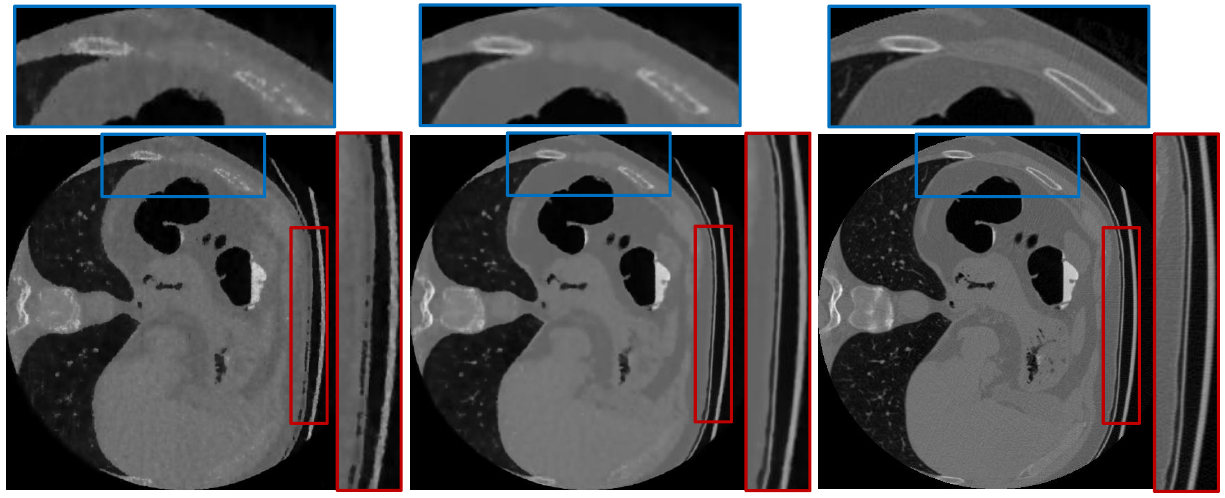}
  % \vspace{1.5cm}
\end{minipage}
\caption{The motivation of frequency regularization in sparse-view tomographic reconstruction. The left and middle columns display the results of NAF and Freq-NAF, respectively, and the right column presents the ground truth.}
\label{fig1}
\end{figure}

Existing methods for tomographic reconstruction can be categorized into three groups: analytical, iterative, and learning-based methods. Analytical methods estimate the attenuation or activity distribution by solving Radon transforms and their inverses. Typical examples include the Feldkamp-Davis-Kress (FDK) algorithm \cite{ref3} and filtered back projection (FBP) \cite{ref4} methods. While these methods yield good results under ideal conditions, they perform poorly when handling ill-posed problems such as sparse-view tomography. The second category involves iterative methods that formulate the reconstruction problem as an optimization process, often incorporating regularization modules. While iterative methods excel in dealing with ill-posed problems \cite{sart,TV,tv-papa}, they require significant computational overhead.

Recently, neural rendering has emerged as an effective technology for synthesizing new views and reconstructing 3D scenes \cite{ref5}. This is made possible by its compact implicit representation. Similar to tomographic image reconstruction, these techniques employ RGB images to learn deep neural radiance fields (NeRF) and offer significant advantages in 3D geometry perception and multi-view consistency.As neural rendering technology rapidly advances, it has been introduced into the reconstruction of CBCT images \cite{ref6,ref7}. Two separate studies have proposed the use of volume rendering for fast CBCT reconstruction. However, both of these studies do not adequately explore strategies for high-quality image reconstruction. Models frequently pay excessive attention to high-frequency information and neglect essential low-frequency information, resulting in inaccurate image reconstruction. This issue is particularly pronounced at the edges and details between slices of images of different organs. Fig.~\ref{fig1}, left column, illustrates the overfitting problem in traditional NAF \cite{ref7} models during abdominal slice reconstruction, resulting in enhanced artifacts in CT images, loss of contrast, spatial resolution.

In response to these observations, we introduce a frequency regularization term \cite{ref8}. This frequency regularization term is employed to stabilize the learning process by regularizing the frequency information of the input images for NAF models. We refer to this approach as \textbf{Freq}uency regularized \textbf{NAF} (Freq-NAF), a self-supervised sparse-view tomographic reconstruction method. As shown in Fig.~\ref{fig1}, middle column, our Freq-NAF model demonstrates superior performance compared to NAF. By introducing frequency regularization, we achieve a better balance between high-frequency and low-frequency information, mitigating overfitting problems. We observe significant improvements in addressing artifacts in CT abdominal slices and noise in SPECT images, along with enhanced spatial resolution and contrast in the reconstructed images.

% We reveal the link between sparse-view tomographic reconstruction and positional encoding frequencies, a discovery which is further corroborated through empirical investigations.
In general, the primary contributions in this work can be summarized as follows: (1) We introduce an entirely novel approach that effectively mitigates overfitting by employing frequency regularization to better balance high-frequency and low-frequency information. It is noteworthy that our method represents the first endeavor to address the issue of sparse-view tomographic reconstruction from a frequency perspective. (2) We conduct comprehensive experimental studies on two prevalent medical imaging modalities, namely CBCT and SPECT. The results unequivocally demonstrate the exceptional performance of our method in various aspects, surpassing recent state-of-the-art methods. This underscores the robustness and broad applicability of our approach.
% , marking significant advancements in the field of sparse-view tomographic reconstruction.

% The rest of this paper is organized as follows. In Section 2, we introduce the details of the proposed Freq-NIF. In Section 3, we present the experimental results on the CBCT and SPECT datasets. Finally, the paper is concluded in Section 4.

% \begin{figure*}[htp!]
% \centering
% \includegraphics[width=.8\textwidth]{figs/freq-NIF最终版.jpg}
% \caption{Freq-NIF pipeline. 
% % a) Tomographic Scanning: Employing a 2D panel for uniformly spaced ray sampling to acquire projection data. 
% a) Ray Sampling: Simulating the accumulation process of intensity values using ray sampling to provide data for subsequent processing.
% b) Position Encoding: Introducing a position encoding network to encode spatial coordinates along the scanning path, extracting valuable features to help the model understand the spatial structure of objects.
% c) Frequency Regularization: Regularizing the visible frequency bands of the position encoding to stabilize the learning process and mitigate early training overfitting, addressing issues such as artifacts and detail loss. d) Intensity Prediction: An MLP network uses encoded information to predict intensity values, a critical step in generating the final image. e) Projection Synthesis: The final step in Freq-NIF involves synthesizing projections by predicting intensity values along incident ray paths, providing data for subsequent image reconstruction.}
% \label{fig2}
% \end{figure*} 

\section{Proposed Method}

\subsection{Positional encoding}

\begin{figure*}[h!]
\centering
\includegraphics[width=.9\textwidth]{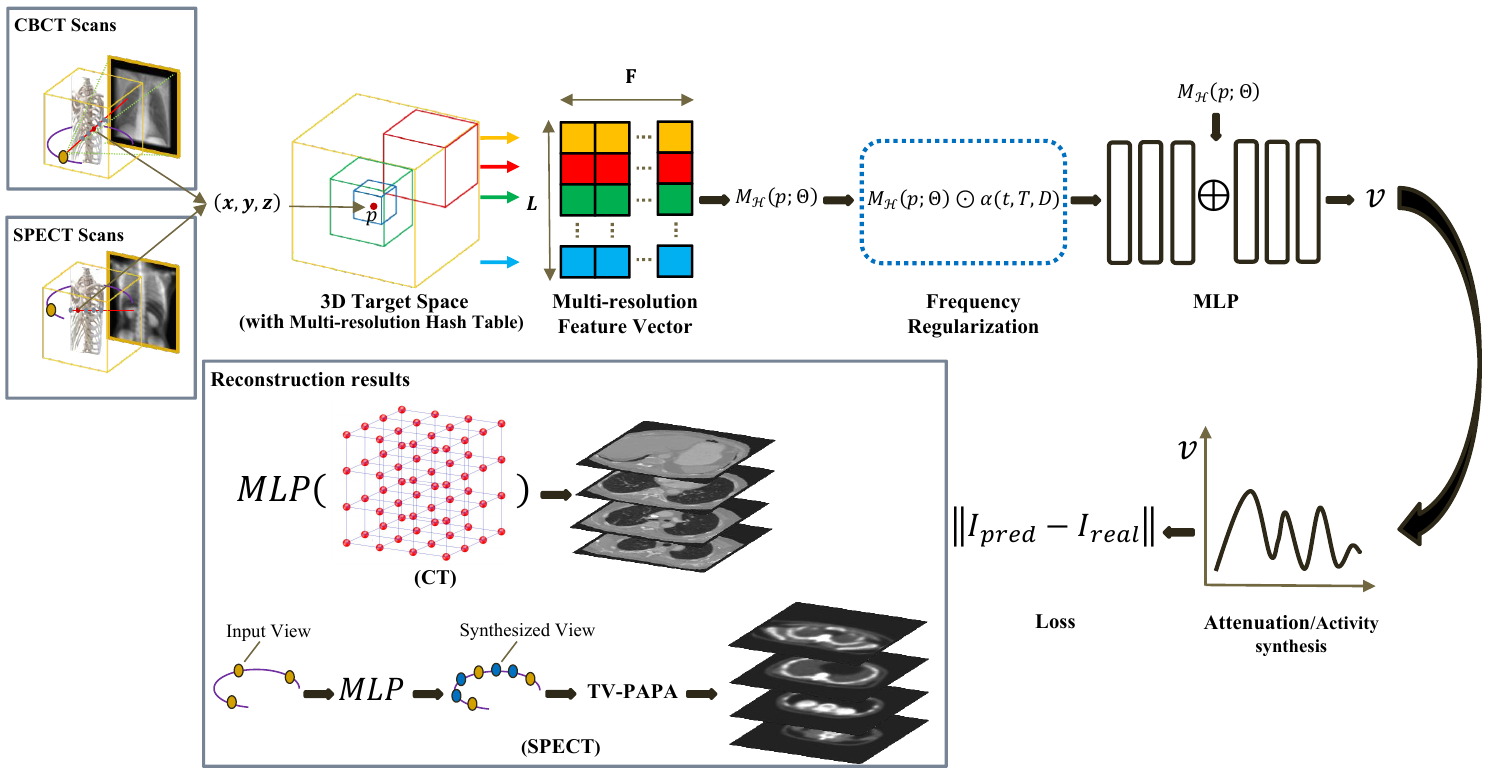}
\caption{ \textbf{Overview of Freq-NAF for sparse-view tomographic reconstruction. }
Given input views, we synthesized projections along X- or gamma-rays with uniformly-sampled points. 
Subsequently, we employed the proposed frequency regularization method to regulate the available frequency bands of positional encoding during training of neural attenuation/activity field. In reconstruction stage, different strategies were employed for CBCT and SPECT modalities. 
% a) Tomographic Scanning: Employing a 2D panel for uniformly spaced ray sampling to acquire projection data. 
% a) Ray Sampling: Simulating the accumulation process of intensity values using ray sampling to provide data for subsequent processing.
% b) Position Encoding: Introducing a position encoding network to encode spatial coordinates along the scanning path, extracting valuable features to help the model understand the spatial structure of objects.
% c) Frequency Regularization: Regularizing the visible frequency bands of the position encoding to stabilize the learning process and mitigate early training overfitting, addressing issues such as artifacts and detail loss. d) Intensity Prediction: An MLP network uses encoded information to predict intensity values, a critical step in generating the final image. e) Projection Synthesis: The final step in Freq-NIF involves synthesizing projections by predicting intensity values along incident ray paths, providing data for subsequent image reconstruction.
}
\label{fig2}
\end{figure*}

The universal approximation theorem \cite{ref12} shows that a pure MLP could approximate any complex functions. However, fitting high-frequency signals with a pure MLP is quite challenging due to the spectral bias problem \cite{ref13, ref14}. To address this issue, we introduce positional encoders that map 3D spatial coordinates into high-dimensional spaces. In Freq-NAF, we utilize the state-of-the-art hash encoders \cite{ref15}, which offer a learning-based sparse coding solution.
\begin{multline}
\small
M_{\mathcal{H}}(p; \Theta) = [\mathcal{I}(H_1), \ldots, \mathcal{I}(H_L)]^{\top}, \\
H = \{c|h(c) = (\boldsymbol{\bigoplus} c_{j} \pi_j) \! \!  \! \! \mod S\}.
\end{multline}
The hash encoder constructs multi-resolution 
voxel grids of $L$ levels, where each voxel grid is associated with a trainable feature lookup table $\Theta$ of size $S$. For each resolution level, we first identify the neighboring corners $c$ of the query point $p$ (cubes of different colors in Fig.~\ref{fig2}), use the hash function $h$ \cite{ref16} to retrieve their corresponding features $H$, and then compute the feature vector through trilinear interpolation $\mathcal{I}$. Finally, concatenating the feature vectors from all resolution levels serves as the output of the hash encoder.

\subsection{Attenuation coefficient/Activity estimation}
We employ a simple MLP $\Phi$ to represent the bounded field, which takes the encoded spatial coordinates as inputs and outputs attenuation coefficient or activity prediction for each position, denoted as $v$.
Specifically, the network consists of 6 fully-connected layers. The first five layers consist of 128 channels, utilizing ReLU activation functions in-between, and the final layer comprises a single neuron with a sigmoid activation function. Additionally, the MLP incorporates skip connections, connecting the network input to the activation of the third layer, to better integrate input information and enhance the accuracy of physical quantity predictions.

\subsection{Attenuation coefficient/Activity synthesis}
% \begin{eqnarray}
% L=L_{a}+L_{s},
% \end{eqnarray}
% Specifically, $L_{a}$ is defined as:
% \begin{eqnarray}
% L_{a}=||p-a||_{1},
% \end{eqnarray}
% where \textit{p }is the prediction result of the QDAN, \textit{a }is the absolute value of $s_{1}$ and $s_{2}$ after subtraction. $s_{1}$ and $s_{2}$ are the subjective scores (i.e., the ground truth) of image $I_{1}$ and image $I_{2}$ respectively.
% The quality assessment loss $L_{s}$ are defined as follows.
% \begin{eqnarray}
% L_{s}=L_{s}^{1}+L_{s}^{2},
% \end{eqnarray}
% \begin{eqnarray}
% L_{s}^{1}=||q_{1}-s_{1}||_{1},
% \end{eqnarray}
% \begin{eqnarray}
% L_{s}^{2}=||q_{2}-s_{2}||_{1},
% \end{eqnarray}
% where $q_{1}$ denotes the prediction result of the main regressor and $q_{2}$ denotes the prediction result of the secondary regressor.
For each 2D scan, the proposed neural attenuation/activity field model takes incident ray intensity and simulates a beam of light passing through different tissues and being detected. According to the Beer-Lambert law, we numerically synthesize the projection process using the following equation:
\begin{equation}
I = I_0 \exp(-\sum_{i=1}^{N} v_i \delta_i)
\label{eq1}.
\end{equation}
Here $I_0$ represents the incident ray intensity, $v_i$ is the underlying physical quantity, that is, attenuation coefficient or activity for CBCT or SPECT, respectively, and $\delta_i = \lVert p_{i+1} - p_i \rVert$ denotes the distance between adjacent sampling points.  

We generate synthesized projection $I_{\text{pred}}$ from the attenuation/activity field using Eq.~\eqref{eq1}, and learn the neural attenuation/activity field via minimizing the $\ell_2$ loss between synthesized and real projections:
\begin{equation}
L(\Theta, \Phi) = \sum_{r \in R} \lVert I_{\text{pred}}(r) - I_{\text{real}}(r) \rVert^2.
\end{equation}
Here $R$ is a ray batch, and $I_{\text{real}}(r)$ is real projection for ray $r$. The hash encoder $\Theta$ and the underlying physical quantity field $\Phi$ are jointly optimized during training process. 
In particular, we employ different strategies to recover tomographic images for different modalities:
\begin{itemize}
     \item For CBCT, we construct a uniform grid on the attenuation field, and discretely sample the trained MLP on the underlying grid to predict the attenuation coefficients.  
    \item For SPECT, we employ the trained MLP to synthesize projection of novel views. Based on the dense combination of training and synthesized projection, we further apply TV-PAPA \cite{tv-papa} to reconstruct SPECT images. 
\end{itemize}

\subsection{Frequency regularization}
In the proposed method, we introduce frequency regularization as a crucial step to mitigate the overfitting issue and improve the quality of sparse-view reconstructed images. It is imperative to clarify the significance of frequency regularization in the context of sparse-view reconstruction. In the following, we briefly elucidate our observations through Fig.~\ref{fig1}, postponing the detailed experimental descriptions to \S\ref{sec:section3}.

In prior study, we have observed that the sparse-view tomographic reconstruction suffers from overfitting, which is embodied in systematic failure in the synthesized-view projection and ray artifacts in the reconstructed image. The recently-developed implicit neural representation (INR) method parameterizes the attenuation coefficient field as a neural network, and performs self-supervised learning via simulating the X-ray attenuation process. As compared to traditional representations (e.g., meshes, point clouds or voxels), INR can better represent complex and highly-detailed distributions of physical quantities and facilitates the learning of hierarchical features of distributions, which ameliorates the issue of overfitting to a certain extent. However, the introduction of position encoder in INR hinders the learning of essential low-frequency information, and misleadingly favors undesired high-frequency artifacts usually present in sparse-view reconstruction. This issue is particularly intense at edges and boundaries in the reconstructed slices, as these regions are frequently degraded by high-frequency artifacts. As depicted in Fig.~\ref{fig1}, we compare the performance of the proposed Freq-NAF model with original NAF on reconstructing abdominal CT image. As shown in the left column of Fig.~\ref{fig1}, we observe that the original NAF is susceptible to overfitting issues. This is evident through noticeable noise in smooth regions, edge damage at bone structures (blue ROI), and loss of contrast (blue ROI) and spatial resolution (red ROI).  

To ameliorate the above issues, we introduce frequency regularization \cite{ref8}, which is a mechanism designed to balance the learning of high-frequency and low-frequency information at different stages of training. The proposed frequency regularization method utilizes a frequency control mechanism: the frequency mask to adjust the visible spectrum based on the time steps of training process. This control mechanism prompts the underlying model to prioritize low-frequency information at the beginning of training, and enables the learning of high-frequency information at subsequent training process after full convergence on low-frequency. In particular, we employ a linearly increasing frequency mask to regulate the available frequency spectrum based on the training steps:
\begin{eqnarray}
{\small
M_D(t, T; p) = M_{\mathcal{H}}(p; \Theta) \odot \alpha(t, T, D) , 
}
\end{eqnarray}
\begin{equation}
{\footnotesize
\begin{aligned}
with \ \alpha_i(t, T, D) =
\begin{cases}
1 & \text{if } i \leq \frac{t \cdot D}{T} + 1, \\
\frac{t \cdot D}{T} - \lfloor\frac{t \cdot D}{T}\rfloor & \text{if } \frac{t \cdot D}{T} + 1 < i \leq \frac{t \cdot D}{T} + 2 ,\\
0 & \text{if } i > \frac{t \cdot D}{T} + 2,
\end{cases}
\end{aligned}
}
\end{equation}
where $\alpha_i(t, T, D)$ represents the $i$-th bit of mask $\alpha(t, T, D)$; $t$ and $T$ correspond to the current and the final iterations for frequency regularization duration, respectively; and $D:=L\times F$ with $F$ denoting the feature dimension per resolution level after hash positional encoding. To be specific, as training proceeds, an increment of 1-bit higher-frequency is augmented as available at each iteration. Frequency regularization gradually introduces higher-frequency information to the neural representation model, which mitigates the impact of unstable high-frequency artifacts or noise at early stage of training, meanwhile minimizing the potential for over-smoothing at the end. 
As shown in the middle column of Fig.~\ref{fig1}, the proposed Freq-NAF exhibits superior performance compared to NAF. By introducing frequency regularization, Freq-NAF achieves a better balance between high-frequency and low-frequency information. Indeed, we observe significant improvements in noise suppression and spatial resolution preservation, as well as contrast enhancement in Freq-NAF restored image. 

\section{Experiments}
\label{sec:section3}

\subsection{Experimental Setups}\label{AA}

\textbf{Data.} We conducted numerical experiments on 4 CBCT datasets and 10 parallel-beam SPECT datasets to demonstrate the effectiveness of the proposed Freq-NAF in sparse-view tomographic reconstruction. Details are listed in Table~\ref{table1}. 

\textbf{CBCT.} We evaluated Freq-NAF using the same publicly available datasets of human organ CT scans as NAF \cite{ref17,ref18}, including chest, jaw, foot and abdomen regions. The chest data was extracted from the LIDC-IDRI dataset \cite{ref17}, while the remaining datasets were obtained from the Open Scientific Visualization Datasets \cite{ref18}. In particular, we generated cone-beam projection data via the application of tomographic toolbox TIGRE \cite{ref19}. We specifically captured 50 projections with $3\%$ Gaussian noise in the range of $180^{\circ}$ under the setting of TIGRE. The proposed Freq-NAF was trained with the above projections, and was evaluated against the raw volumetric CT data. 

\textbf{SPECT.} The SPECT projection datasets were acquired from XCAT \cite{ref20} anthropomorphic digital phantoms using Monte Carlo simulation software SIMIND \cite{ref21,ref22}. 
In particular, we simulated 10 XCAT phantoms for the bone SPECT study with radiotracer Tc-99m-MDP. A SIEMENS E.CAM gamma camera equipped with a low energy high resolution parallel-beam collimator was simulated. The detector orbit was circular covering $360^{\circ}$, and a $10\%$ detector energy
resolution centered at 140 keV was employed for all SPECT simulations. The sparse-view projection data consisted of 30 SIMIND-simulated projections (referred to as real projections), which were used for self-supervised training to determine the INR model. We then synthesized 90 novel-view projections based on the optimized INR model, and employed TV-PAPA to reconstruct the combination of 120 projections.

\begin{table}[t]
\centering
\caption{Details of datasets used in the experiments.}
\label{table1}
\resizebox{\linewidth}{!}{%
\begin{tabular}{ccccccc}
\toprule
Dataset Name & \thead{Dimension} & \thead{Scanning \\ Method} & \thead{Scanning \\ Range} & \thead{Number of \\ Projections} & \thead{Detector \\ Resolution} \\
\midrule
Chest (CT)\cite{ref17} & 128$\times$128$\times$128 & TIGRE\cite{ref19} & 0\textdegree~$\sim$ 180\textdegree & 50 & 256$\times$256 \\
Jaw (CT)\cite{ref18} & 256$\times$256$\times$256 & TIGRE\cite{ref19} & 0\textdegree~$\sim$ 180\textdegree & 50 & 512$\times$512 \\
Foot (CT)\cite{ref18} & 256$\times$256$\times$256 & TIGRE\cite{ref19} & 0\textdegree~$\sim$ 180\textdegree & 50 & 512$\times$512 \\
Abdomen (CT)\cite{ref18} & 512$\times$512$\times$463 & TIGRE\cite{ref19} & 0\textdegree~$\sim$ 180\textdegree & 50 & 1024$\times$1024 \\
Torso (SPECT)\cite{ref20} & 128$\times$128$\times$128 & SIMIND\cite{ref21} & 0\textdegree~$\sim$ 360\textdegree & 30 & 128$\times$128 \\
\bottomrule
\end{tabular}
}
\end{table}

\textbf{Baseline methods.} To validate the superiority of our method, we compare Freq-NAF against both learning-based and handcrafted baseline methods. For CBCT, we choose FDK\cite{ref3} as a representative analytical method and SART\cite{sart} as a robust iterative reconstruction algorithm, along with the NeRF-based method NAF\cite{ref7}. For SPECT, we choose traditional techniques including FBP\cite{ref4} and TV-PAPA\cite{tv-papa}, alongside the NeRF-based method NAF\cite{ref7}.

\textbf{Implementation details.} We implement Freq-NAF using PyTorch with a single NVIDIA RTX 3090Ti GPU. We use the Adam optimizer with a learning rate that starts at $1 \times 10^{-3}$ and steps down to $1 \times 10^{-4}$.
% For chest, abdomen CT scans, and torso SPECT, our model undergoes 1500 epochs of training. In contrast, jaw and foot CT datasets require 3000 epochs for comprehensive training.
The batch size is 1024 rays at each iteration. The sampling quantity of each ray depends on the size of the tomographic image. For example, we sample 320 points along each ray for the $ 256 \times 256 \times 256 $ foot CT.
Our implementation follows the official methodology outlined in \cite{ref15} for the hash table encoder. We fine-tune the core parameters to obtain optimal performance. This includes setting $S$ (the size of hash table) to $2^{19}$, $L$ (the number of resolution levels) to 16, and $F$ (the feature dimension) to 8. We further set the end iteration for frequency regularization as $T = \lfloor x\% \times \text{total\_iters} \rfloor$.

\begin{figure*}[htp!]
\centering
\includegraphics[width=.95\textwidth ]{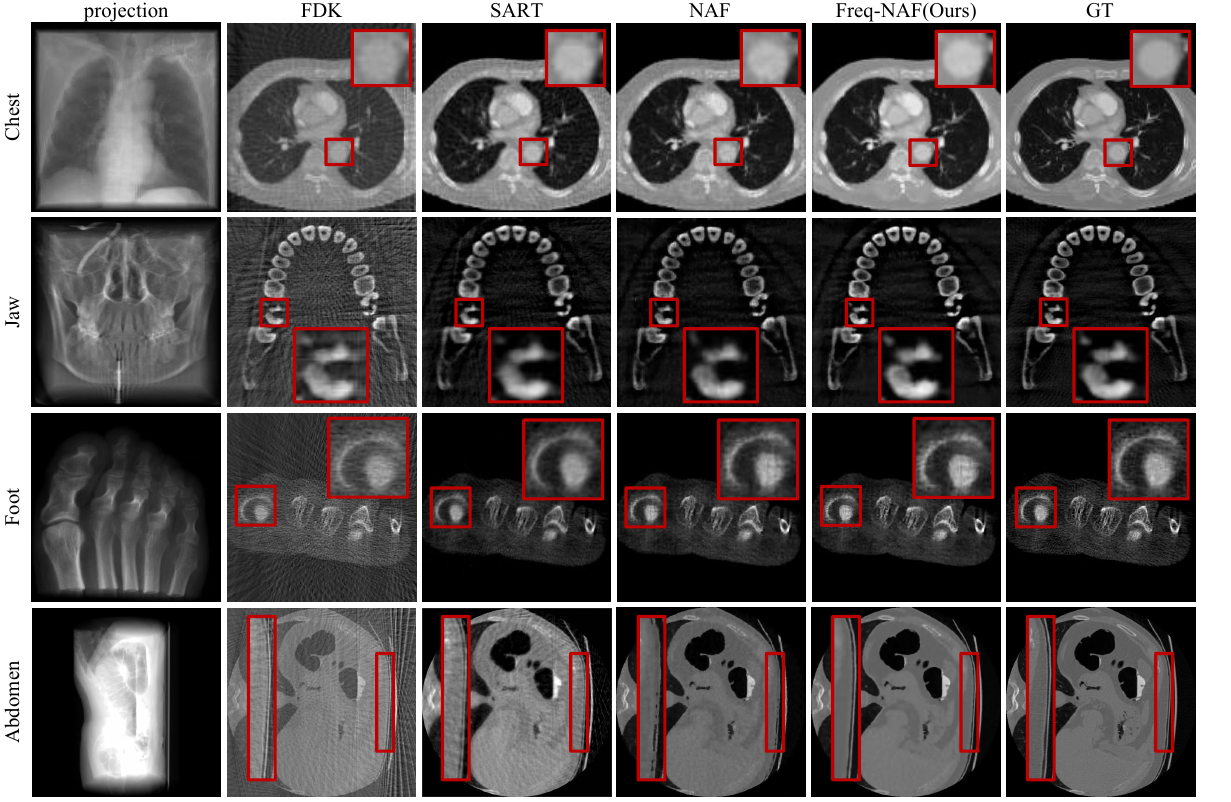}
\caption{ Qualitative results of four methods. From left to right: examples of Xray projections, slices of 3D CT models reconstructed by four methods, and the ground truth CT slices. Please zoom in for a better view.}
\label{fig3}
\end{figure*}
\begin{figure*}[h]
\centering
\includegraphics[width=.95\textwidth]{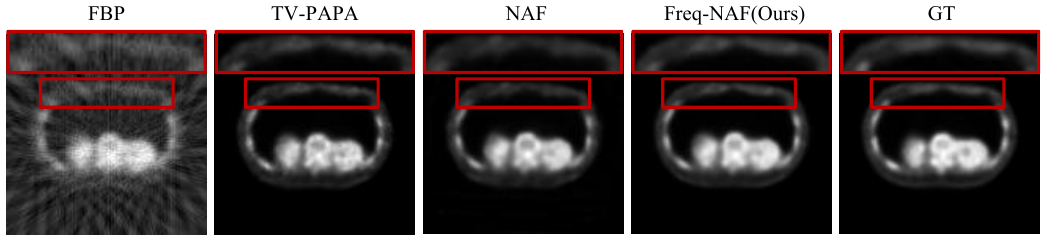}
\caption{ Qualitative result of SPECT reconstruction. Please zoom in for a better view.}
% \caption{ Qualitative results of four methods. From left to right: slices of 3D SPECT models reconstructed by four methods, and the ground truth SPECT slices. Please zoom in for a better view.}
\label{fig4}
\end{figure*}

\textbf{Evaluation metrics.} To quantitatively evaluate the performance of the four methods, we adopt three image quality metrics. We first compute PSNR to evaluate pseudo-shadow suppression capability and SSIM \cite{ref23} to measure the perceptual differences between pairs of signals. Additionally, we calculate Learned Perceptual Image Patch Similarity (LPIPS) \cite{ref24}, a learning-based perceptual similarity metric, for quantitative evaluation. 

\subsection{Experimental Results}
\textbf{CBCT.} In Table~\ref{tab:performance}, we present quantitative performance metrics for various methods on human organ. Our results reveal that the proposed Freq-NAF method consistently outperforms other techniques across all evaluated organs, exhibiting superior PSNR and SSIM scores while maintaining lower LPIPS values. For example, the PSNR value of our method in the abdomen dataset is 2.0 dB higher than that of the second-best method NAF. Similarly impressive results were observed for the chest, jaw, and foot organs, as outlined in Table~\ref{tab:performance}.

In Fig.~\ref{fig3}, we provide visualization results of different methods. FDK produces results with numerous streak artifacts due to the limited number of projection views. SART generates results with well-defined shapes but lacks detailed internal information. For NAF, the edges between materials appear slightly blurred, indicating that the frequency encoder fails to teach the network to focus on edges. However, with the assistance of frequency regularization, Freq-NAF exhibits the richest details, clearest edges and least artifacts.
\begin{table}[htp!]
\centering
\caption{ PSNR↑ / SSIM↑ / LPIPS↓ measurements of four methods on four CBCT datasets.}
\label{tab:performance}
\resizebox{\linewidth}{!}{%
\begin{tabular}{ccccc}
\toprule
Method & Chest & Jaw & Foot & Abdomen \\
\midrule
FDK\cite{ref3} & 22.89/.782/.261 & 28.59/.780/.377 & 23.92/.580/.591 & 22.39/.586/.476 \\
SART\cite{sart} & 32.12/.948/.070 & 32.68/.932/\textbf{.118} & 30.13/.930/\textbf{.053} & 31.38/.918/.229 \\
NAF\cite{ref7} & 33.07/.963/.051 & 34.47/.940/.142 & 31.67/.935/.106 & 34.45/.955/.199 \\
Freq-NAF (Ours) & \textbf{34.18}/\textbf{.973}/\textbf{.024} & \textbf{36.25}/\textbf{.958}/.124 & \textbf{32.39}/\textbf{.941}/.090 & \textbf{36.45}/\textbf{.967}/\textbf{.188} \\
\bottomrule
\end{tabular}
}
\end{table}

% Fig.~\ref{fig4} illustrates the performance of iterative and learning-based methods under varying numbers of input views. It is evident that performance improves with an increase in the number of input views. In most cases, our method outperforms others.

\textbf{SPECT.} We select 300 abdomen images from 10 XCAT phantoms to showcase the performance of various methods. In Table~\ref{table3}, Freq-NAF produces best quantitative results, with significantly better PSNR, SSIM and LPIPS values compared to other methods. Fig.~\ref{fig4} displays the visual outcomes of various methods. Our method exhibits higher contrast and lower noise levels compared to FBP, TV-PAPA and NAF.
\begin{table}
\centering
\caption{ PSNR↑/SSIM↑/LPIPS↓ measurements of four methods 
 on Abdomen SPECT Dataset.}
\begin{tabular}{lccc}
\toprule
Method & PSNR↑ & SSIM↑ & LPIPS↓ \\
\midrule
FBP\cite{ref4} & 18.15 & 0.189 & 0.391 \\
TV-PAPA\cite{tv-papa} & 32.39 & 0.960 & 0.238 \\
NAF\cite{ref7} & 32.10 & 0.902 & 0.011 \\
Freq-NAF (Ours) & \textbf{33.09} & \textbf{0.971} & \textbf{0.007} \\
\bottomrule
\end{tabular}
\label{table3}
\end{table}

\subsection{Ablation Study}
\begin{figure}[htp!]
\centering
\includegraphics[width=1.\columnwidth]{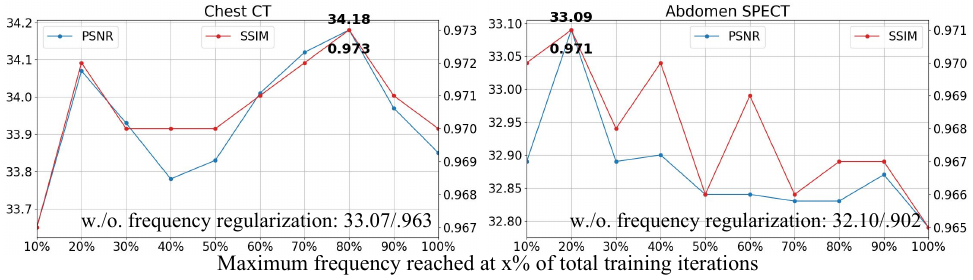}
\caption{ Effects of frequency regularization duration on CT and SPECT datasets. We set the end iteration for frequency regularization as $T = \lfloor x\% \times \text{total\_iters} \rfloor$. Freq-NAF achieves good performance across a wide range of duration choices, which uniformly outperform ``w/o. frequency regularization". }
\label{fig5}
\end{figure}

In this section, we evaluate the performance of frequency regularization duration $T$ under different abundance of high-frequency components. Indeed, the optimal frequency regularization duration increases with the abundance of high-frequency components in the underlying dataset. For instance, the abdomen CT dataset contains a substantial amount of high-frequency details. In this case, a longer frequency regularization duration is required to achieve the optimal performance, where the $100\%$-schedule performs best. As the chest and jaw CT datasets still exhibit a certain amount of high-frequency structures, the respective optimal durations are empirically determined as $80\%$ and $40\%$ of total iterations. As a contrast, the XCAT SPECT dataset exhibits a smooth distribution, which naturally leads to a lower $20\%$ duration. In particular, we show the effects of frequency regularization duration on chest CT and abdomen SPECT datasets in Fig.~\ref{fig5}.

\section{Conclusion}
This work deals with the challenge in sparse-view tomographic reconstruction by proposing Freq-NAF, a self-supervised method that improves the neural attenuation/activity field through the incorporation of frequency regularization. The introduction of frequency regularization effectively balances high-frequency and low-frequency information, alleviating the overfitting issue. Through numerical experiments conducted on CBCT and SPECT datasets, we demonstrate the state-of-the-art accuracy achieved by our method. We believe that our work offers new perspectives and promising directions for research in the realm of high-quality, low-dose sparse-view tomographic reconstruction.
\bibliographystyle{IEEEbib}
\bibliography{icme2023template_new}

\begin{thebibliography}{10}

\bibitem{ref1}
Paul Suetens,
\newblock {\em Fundamentals of medical imaging},
\newblock Cambridge university press, 2017.

\bibitem{ref2}
Saiprasad Ravishankar, Jong~Chul Ye, and Jeffrey~A Fessler,
\newblock ``Image reconstruction: From sparsity to data-adaptive methods and machine learning,''
\newblock {\em Proceedings of the IEEE}, vol. 108, no. 1, pp. 86--109, 2019.

\bibitem{ref3}
Lee~A Feldkamp, Lloyd~C Davis, and James~W Kress,
\newblock ``Practical cone-beam algorithm,''
\newblock {\em Josa a}, vol. 1, no. 6, pp. 612--619, 1984.

\bibitem{ref4}
Matteo Ronchetti,
\newblock ``Torchradon: Fast differentiable routines for computed tomography,''
\newblock {\em arXiv preprint arXiv:2009.14788}, 2020.

\bibitem{sart}
Anders~H Andersen and Avinash~C Kak,
\newblock ``Simultaneous algebraic reconstruction technique (sart): a superior implementation of the art algorithm,''
\newblock {\em Ultrasonic imaging}, vol. 6, no. 1, pp. 81--94, 1984.

\bibitem{TV}
Emil~Y Sidky and Xiaochuan Pan,
\newblock ``Image reconstruction in circular cone-beam computed tomography by constrained, total-variation minimization,''
\newblock {\em Physics in Medicine \& Biology}, vol. 53, no. 17, pp. 4777, 2008.

\bibitem{tv-papa}
Si~Li, Jiahan Zhang, Andrzej Krol, C~Ross Schmidtlein, Levon Vogelsang, Lixin Shen, Edward Lipson, David Feiglin, and Yuesheng Xu,
\newblock ``Effective noise-suppressed and artifact-reduced reconstruction of spect data using a preconditioned alternating projection algorithm,''
\newblock {\em Medical physics}, vol. 42, no. 8, pp. 4872--4887, 2015.

\bibitem{ref5}
Ben Mildenhall, Pratul~P Srinivasan, Matthew Tancik, Jonathan~T Barron, Ravi Ramamoorthi, and Ren Ng,
\newblock ``Nerf: Representing scenes as neural radiance fields for view synthesis,''
\newblock {\em Communications of the ACM}, vol. 65, no. 1, pp. 99--106, 2021.

\bibitem{ref6}
Darius R{\"u}ckert, Yuanhao Wang, Rui Li, Ramzi Idoughi, and Wolfgang Heidrich,
\newblock ``Neat: Neural adaptive tomography,''
\newblock {\em ACM Transactions on Graphics (TOG)}, vol. 41, no. 4, pp. 1--13, 2022.

\bibitem{ref7}
Ruyi Zha, Yanhao Zhang, and Hongdong Li,
\newblock ``Naf: Neural attenuation fields for sparse-view cbct reconstruction,''
\newblock in {\em International Conference on Medical Image Computing and Computer-Assisted Intervention}. Springer, 2022, pp. 442--452.

\bibitem{ref8}
Jiawei Yang, Marco Pavone, and Yue Wang,
\newblock ``Freenerf: Improving few-shot neural rendering with free frequency regularization,''
\newblock in {\em Proceedings of the IEEE/CVF Conference on Computer Vision and Pattern Recognition}, 2023, pp. 8254--8263.

\bibitem{ref12}
Kurt Hornik, Maxwell Stinchcombe, and Halbert White,
\newblock ``Multilayer feedforward networks are universal approximators,''
\newblock {\em Neural networks}, vol. 2, no. 5, pp. 359--366, 1989.

\bibitem{ref13}
Nasim Rahaman, Aristide Baratin, Devansh Arpit, Felix Draxler, Min Lin, Fred Hamprecht, Yoshua Bengio, and Aaron Courville,
\newblock ``On the spectral bias of neural networks,''
\newblock in {\em International conference on machine learning}. PMLR, 2019, pp. 5301--5310.

\bibitem{ref14}
Zhi-Qin~John Xu, Yaoyu Zhang, Tao Luo, Yanyang Xiao, and Zheng Ma,
\newblock ``Frequency principle: Fourier analysis sheds light on deep neural networks,''
\newblock {\em arXiv preprint arXiv:1901.06523}, 2019.

\bibitem{ref15}
Thomas M{\"u}ller, Alex Evans, Christoph Schied, and Alexander Keller,
\newblock ``Instant neural graphics primitives with a multiresolution hash encoding,''
\newblock {\em ACM transactions on graphics (TOG)}, vol. 41, no. 4, pp. 1--15, 2022.

\bibitem{ref16}
Matthias Teschner, Bruno Heidelberger, Matthias M{\"u}ller, Danat Pomerantes, and Markus~H Gross,
\newblock ``Optimized spatial hashing for collision detection of deformable objects.,''
\newblock in {\em Vmv}, 2003, vol.~3, pp. 47--54.

\bibitem{ref17}
Samuel~G Armato~III, Geoffrey McLennan, Luc Bidaut, Michael~F McNitt-Gray, Charles~R Meyer, Anthony~P Reeves, Binsheng Zhao, Denise~R Aberle, Claudia~I Henschke, Eric~A Hoffman, et~al.,
\newblock ``The lung image database consortium (lidc) and image database resource initiative (idri): a completed reference database of lung nodules on ct scans,''
\newblock {\em Medical physics}, vol. 38, no. 2, pp. 915--931, 2011.

\bibitem{ref18}
P.~Klacansky,
\newblock ``Open scientific visualization datasets,'' 2022.

\bibitem{ref19}
Ander Biguri, Manjit Dosanjh, Steven Hancock, and Manuchehr Soleimani,
\newblock ``Tigre: a matlab-gpu toolbox for cbct image reconstruction,''
\newblock {\em Biomedical Physics \& Engineering Express}, vol. 2, no. 5, pp. 055010, 2016.

\bibitem{ref20}
W~Paul Segars, G~Sturgeon, S~Mendonca, Jason Grimes, and Benjamin~MW Tsui,
\newblock ``4d xcat phantom for multimodality imaging research,''
\newblock {\em Medical physics}, vol. 37, no. 9, pp. 4902--4915, 2010.

\bibitem{ref21}
Michaella Morphis, Johan~A van Staden, Hanlie du~Raan, and Michael Ljungberg,
\newblock ``Modelling of energy-dependent spectral resolution for spect monte carlo simulations using simind,''
\newblock {\em Heliyon}, vol. 7, no. 2, 2021.

\bibitem{ref22}
Michael Ljungberg, Sven-Erik Strand, and Michael~A King,
\newblock {\em Monte Carlo calculations in nuclear medicine: Applications in diagnostic imaging},
\newblock CRC Press, 2012.

\bibitem{ref23}
Zhou Wang, Alan~C Bovik, Hamid~R Sheikh, and Eero~P Simoncelli,
\newblock ``Image quality assessment: from error visibility to structural similarity,''
\newblock {\em IEEE transactions on image processing}, vol. 13, no. 4, pp. 600--612, 2004.

\bibitem{ref24}
Richard Zhang, Phillip Isola, Alexei~A Efros, Eli Shechtman, and Oliver Wang,
\newblock ``The unreasonable effectiveness of deep features as a perceptual metric,''
\newblock in {\em Proceedings of the IEEE conference on computer vision and pattern recognition}, 2018, pp. 586--595.

\end{thebibliography}
\vspace{12pt}
\end{document}